\documentclass[journal]{IEEEtran}
\usepackage{fancyhdr}
\usepackage{amsmath,amssymb}
\usepackage{amsfonts}
\usepackage[dvips]{graphicx}
\usepackage{dsfont}
\usepackage{balance}
\usepackage{algpseudocode}
\usepackage{algorithm}

\usepackage[hidelinks]{hyperref}    

\usepackage{color}
\usepackage{pgfplots}           
\pgfplotsset{compat=1.16}

\usepackage{enumerate}
\usepackage{graphics}
\usepackage{subfigure}
\usepackage{cite}
\usepackage{mathrsfs}
\usepackage{stfloats}

\usepackage{tikz}
\usepackage{mathdots}
\usepackage{yhmath}
\usepackage{cancel}
\usepackage{siunitx}
\usepackage{array}
\usepackage{multirow}
\usepackage{textcomp}
\usepackage{gensymb}
\usepackage{tabularx}
\usepackage{booktabs}
\usepackage{graphicx}

\usepackage{enumitem}

\newtheorem{theorem}{Theorem}

\newtheorem{corollary}{Corollary}

\newtheorem{proposition}{Proposition}
\newtheorem{remark}{Remark}

\begin{document}

\title{On the Distribution of the Sum of Double-Nakagami-$m$ Random Vectors and Application in Randomly Reconfigurable Surfaces}
\author{
Sotiris A. Tegos,~\IEEEmembership{Student Member,~IEEE,}
Dimitrios Tyrovolas,~\IEEEmembership{Student Member,~IEEE,}\\
Panagiotis D. Diamantoulakis,~\IEEEmembership{Senior Member,~IEEE,}
Christos K. Liaskos,~\IEEEmembership{Member,~IEEE,} \\
and George K. Karagiannidis,~\IEEEmembership{Fellow,~IEEE}
\thanks{S. A. Tegos, D. Tyrovolas, P. D. Diamantoulakis and G. K. Karagiannidis are with the Wireless Communications and Information Processing (WCIP) Group, Electrical \& Computer Engineering Dept., Aristotle University of Thessaloniki, 54 124, Thessaloniki, Greece (e-mails: \{tegosoti,tyrovolas,padiaman,geokarag\}@auth.gr).
C. K. Liaskos is with the Computer Science Engineering Department, University of Ioannina,
45110 Ioannina, Greece (e-mail: cliaskos@cse.uoi.gr).}
}
\maketitle
%

\begin{abstract} 
Meta-surfaces intend to improve significantly the performance of future wireless networks by controlling the wireless propagation and shaping the radio waves according to the generalized Snell’s law. 
A recent application of meta-surfaces is reconfigurable intelligent surfaces which are practically limited by the requirement for perfect knowledge of the user's position. For the case where the user's position cannot be obtained, we introduce randomly reconfigurable surfaces (RRSs) aiming to diffuse the incoming wave. A RRS is defined as a reconfigurable meta-surface that each of its elements induces a randomly selected time-variant phase shift on the reflected signal. 
To facilitate the performance analysis of a RRS-assisted system, first, we present novel closed-form expressions for the probability density function, the cumulative distribution function, the moments, and the characteristic function of the distribution of the sum of double-Nakagami-$m$ random vectors, whose amplitudes follow the double-Nakagami-$m$ distribution, i.e., the distribution of the product of two random variables following the Nakagami-$m$ distribution, and phases follow the circular uniform distribution.
We also consider a special case of this distribution, namely the distribution of the sum of Rayleigh-Nakagami-$m$ random vectors. 
Then, we exploit these expressions to investigate the performance of the RRS-assisted composite channel, assuming that the two links undergo Nakagami-$m$ fading and the equivalent phase follows the circular uniform distribution.
Closed-form expressions for the outage probability, the average received signal-to-noise ratio, the ergodic capacity, the bit error probability, the amount of fading, and the channel quality estimation index are provided to evaluate the performance of the considered system. These metrics are also derived for the practical special case where one of the two links undergoes Rayleigh fading.
\end{abstract}



\section{Introduction}

\IEEEPARstart{N}{owadays}, the need for higher data rate grows exponentially as the number of smart devices demanding ultra reliable and low latency connection to mobile networks keeps increasing following Edholm's law \cite{Cherry2004}. Fifth generation (5G) wireless networks enable massive communications and are able to serve the users with a wide range of services such as autonomous vehicles, virtual reality applications, industrial internet of things (IIoT), etc. 

The upcoming advent of the sixth generation (6G) wireless networks requires ultra low latency, coverage extension,  improved throughput and security leading to communications on higher frequency bands like millimeter wave (mmWave) and wireless optical communications \cite{Pan2020, Yadav2018}. 
The main obstacles of communications in these bands are the sensitivity to blockage due to higher atmospheric attenuation, thus severe path loss and limited coverage. 
One solution for the aforementioned problem is the denser deployment of base stations (BSs) which can fill coverage holes, reduce the path loss, and facilitate higher rates, but with a higher cost and power consumption.
To this end, the development of a smart radio environment (SRE) \cite{Letaief2019} is considered as an operational alternative due to their ability to provide large spectrum bandwidth with lower cost compared to the denser deployment of BSs.

\subsection{Motivation}

Until recently, the propagation medium in the field of wireless communications has been perceived as a randomly behaving entity between the communicating systems, which negatively affects the quality of the received signal, due to the uncontrollable interactions (reflections, scattering, etc.) of the transmitted radio waves with the surrounding objects \cite{Basar2019}. However, in the future SREs, the wireless propagation is envisioned to be controllable and the information to be processed through the transmission. This is facilitated by the ability of SREs to provide future wireless networks with uninterrupted wireless connectivity using meta-surfaces. 

A meta-surface is an artificial structure with engineered properties, able to alter any impinging electromagnetic wave according to its frequency in a desired way \cite{Yang2016, Liaskos2018}. Practically, a meta-surface is a surface with small but not-negligible depth that consists of a two-dimensional array of sub-wavelength metallic or dielectric scattering elements. Specifically, by changing the scattering elements impedance, the incident wave can be transformed into specified waves. The difference between a conventional surface, e.g., metallic plate, and a meta-surface is the capability of the latter to shape the radio waves according to the generalized Snell's law of reflection and refraction, i.e., the angle of the incident and the reflected wave is not the same \cite{DiRenzo2019}. However, to fulfill the vision of the SREs, the properties of a meta-surface should be readjustable to provide to the users the required quality of service (QoS).

Reconfigurable intelligent surfaces (RISs) have been introduced as meta-surfaces connected with a controller whose properties can be real-time altered according to the controller's signals and, thus, adjust to the network demands. In fact, each RIS incorporates a lightweight gateway, which enables it to receive commands and in collaboration with existing beamforming and localization mechanisms, its properties are adjusted by taking into account the users' location and demanded services. By choosing properly the properties of the RIS, it can implement a plethora of electromagnetic functions such as reflection, steering, diffusion, absorption, etc \cite{Liaskos2018}. However, in most of the existing studies, RISs utilize the steering function, inducing real-time adjustable phase shifts on the reflected signals aiming to focus the signal on a specific user. 

In most of the examined RIS-assisted systems, it is assumed that the location of the user is perfectly known which can be achieved either with localization mechanisms or with perfect channel state information (CSI) acquisition. However, due to the fact that RISs are considered as nearly passive circuits, channel estimation units cannot be integrated into them, thus in practical RIS-assisted systems perfect phase estimation could be difficult. 
In \cite{Badiu2020}, the communication through a RIS whose phase errors follow the Von Mises distribution, was investigated. This setup could represent imperfect phase estimation, quantized reflection phases, or both of them. The composite channel was proved to be equivalent to a point-to-point Nakagami-$m$ fading channel by using the central limit theorem (CLT). However, the CLT does not provide accurate results for a RIS with small number of elements. 

Regarding the fading model, Rayleigh fading model is frequently considered for performance analysis \cite{Basar2019, Nadeem2020, Zhang2019, Qian2020} which is not a legit choice for a practical RIS-assisted communication scenario due to the fact that RISs are carefully deployed to leverage line-of-sight (LoS) links between the terminals.
In \cite{Trigui2020}, a comprehensive theoretical framework for the performance characterization of RIS-assisted communications in different propagation scenarios was provided. Closed-form expressions for the outage probability and the ergodic capacity for different fading environments, were derived assuming perfect phase estimation. 

Concluding the state-of-the-art, most of the recent works focus on RIS-assisted systems without taking into consideration the practical limitations of RISs, i.e., the requirement of perfect CSI or perfect localization. In fact, if the user's location cannot be acquired, it would not be optimal to steer the incoming wave to a random point. A possible strategy for such scenarios could be to diffuse the incoming electromagnetic waves in the half-space in front of the RIS which can be achieved by changing the radiation pattern of the RIS randomly in every coherence time. To the best of the authors' knowledge, the performance of a wireless communication network assisted by a RIS, which utilizes the diffusion function, has not been investigated in the existing literature. In fact, such a system does not require any location or channel state information. Besides, this network can be useful in multi-user systems where the perfect phase adjustment of RISs may be difficult or even impossible, especially as the number of users increases. For example, perfect phase adjustment is impractical in machine-type communication scenarios, where a large number of randomly deployed nodes that occasionally transmit short messages should be served with a random access protocol. It should be highlighted that if a non-reconfigurable meta-surface with fixed phase shifts is deployed in a Rayleigh fading environment, where the phase follows the circular uniform distribution, the same result can be achieved. However, if a meta-surface with fixed phase shifts is used in environments with different fading models, e.g., Nakagami fading, where the phase does not follow the circular uniform distribution, the signals are not diffused. To this end, by taking into account that a RIS could not be placed in a location where the BS-RIS and RIS-user links are characterized by Rayleigh fading, if the location of the user cannot be acquired, a reconfigurable meta-surface which induces random phase shifts should be utilized to diffuse the impinging signal.

\subsection{Contribution}

In this paper, we introduce randomly reconfigurable surfaces (RRSs), being defined as reconfigurable meta-surfaces whose elements induce a randomly selected time-variant phase shift on the reflected signal, in order to diffuse the impinging signal. 
The proposed blind diffusion manipulation of the impinging waves is intended as an additional capability of a software-programmable meta-surface, when the positions of the transmitter or the receiver are unknown or CSI derivation loops are not possible. 
To investigate a RRS-assisted system, we first utilize the distribution of the sum of double-Nakagami-$m$ random vectors, and provide exact closed-form expressions for its statistical properties and the performance metrics of the considered system.
Specifically, the contributions of this work are listed below:
\begin{itemize}
	\item We investigate the distribution of the sum of double-Nakagami-$m$ random vectors, whose amplitudes follow the double-Nakagami-$m$ distribution, i.e., the distribution of the product of two random variables (RVs) following the Nakagami-$m$ distribution, and phases are circular uniformly distributed, and provide exact closed form expressions for its statistical properties, i.e., the PDF, the cumulative distribution function (CDF), the moments and the characteristic function. These statistical properties are also derived for the distribution of the sum of Rayleigh-Nakagami-$m$ random vectors which is a special case of the former distribution, where the amplitudes of the vectors follow the distribution of the product of a RV following the Rayleigh distribution and one following the Nakagami-$m$ distribution.
	
	\item We utilize the derived expressions to investigate a practical RRS-assisted system where Nakagami-$m$ fading channel is assumed both between the BS and the RRS and between the RRS and the user. 
	We extract useful metrics to investigate the performance of the considered system, such as the outage probability, the average received signal-to-noise ratio (SNR), which is proved to be proportional to the number of elements, the ergodic capacity, the bit error probability (BEP) for both binary and $M$-ary modulations, the amount of fading (AoF) and the channel quality estimation index (CQEI). These metrics are also provided for the practical special case where one of the links undergoes Rayleigh fading, where the derived expressions are significantly simplified. Also, it should be highlighted that the performance achieved in the RRS-assisted system can be considered as a lower bound of the performance of a RIS-assisted system.
	
\end{itemize}

\subsection{Structure}

The rest of the paper is organized as follows:
in Section II, the statistical properties of the distribution of the sum of double-Nakagami-$m$ random vectors and the distribution of the sum of Rayleigh-Nakagami-$m$ random vectors are provided.
In Section III, a RRS-assisted system is investigated, useful performance metrics are provided and numerical results are provided to illustrate the performance of the considered system.
Finally, closing remarks and discussion are provided in Section IV.

\section{The Distribution of the Sum of Double-Nakagami-$m$ Random Vectors}

We consider $N$ double-Nakagami-$m$ random vectors, $X_k$, $k\in\{1,N\}$, with amplitudes $h_k$ and phases $\theta_k$, i.e.,
\begin{equation}
	X_k = h_k e^{j \theta_k}.
\end{equation}
The amplitudes $h_k$ are independent and identically distributed (i.i.d.) RVs, which follow the double-Nakagami-$m$ distribution with parameters $m_1$, $m_2$, $\Omega_1$ and $\Omega_2$, i.e., the distribution of the product of two RVs following the Nakagami-$m$ distribution, whose PDF is given by \cite{Karagiannidis2007}
\begin{equation} \label{dNak_PDF}
	\begin{split}
		f_{h_k} (z) & = \frac{4 z^{m_1+m_2-1}}{\Gamma(m_1)\Gamma(m_2)}\left( \frac{m_1 m_2}{\Omega_1 \Omega_2} \right)^\frac{m_1+m_2}{2} \\ 	
			& \times K_{m_1-m_2}\left(2z \sqrt{\frac{m_1 m_2}{\Omega_1 \Omega_2}}\right),
	\end{split}
\end{equation}
where $\Gamma(\cdot)$ is the Gamma function \cite{Gradshteyn2014}.
Moreover, the phases $\theta_k$ are uniformly distributed in $[0,2\pi]$. 

Next, we define the vector $H$ as
\begin{equation}
	H = \sum_{k=1}^{N} X_k.
\end{equation}
\begin{theorem} \label{theorem}
	The PDF of $|H|$ can be expressed in closed-form as
	\begin{equation} \label{PDF}
	\begin{split}
	f_{|H|}(r) & = \sum_{k_1=0}^{m_1-1} ... \sum_{k_N=0}^{m_1-1} 
	\prod_{i=1}^{N} \frac{(m_2)_{m_1-1-k_i}(1-m_2)_{k_i}}{(m_1-1-k)! k_i!} \\
	& \times \frac{4 \left(\frac{m_1 m_2}{\Omega_1 \Omega_2}\right)^{\frac{u+1}{2}}}{(u-1)!} r^u K_{u-1} \left(2 \sqrt{\frac{m_1 m_2}{\Omega_1 \Omega_2}}r\right)\!,
	\end{split}
	\end{equation}
\end{theorem}
where $u = N(m_1+m_2-1) - \sum_{i=1}^{N} k_i$, $K_v()$ is the $v$-th order modified Bessel function of the second kind \cite{Gradshteyn2014}, $(n)_k$ is the Pochhammer symbol and $k!$ is the factorial of $k$. 
\begin{IEEEproof}
	The proof is provided in Appendix \ref{proof_theorem}.
\end{IEEEproof}
It should be highlighted that $m_1$ and $m_2$ are used interchangeably.
In the following proposition, we present the PDF for a special case, where the amplitudes of the vectors follow the distribution of the product of a RV following the Rayleigh distribution and one following the Nakagami-$m$ distribution, i.e., the distribution of the sum of Rayleigh-Nakagami-$m$ random vectors.
\begin{proposition} 
	The PDF of $|H|$, when $H$ follows the distribution of the sum of Rayleigh-Nakagami-$m$ random vectors, can be expressed in closed-form as
	\begin{equation} \label{PDF_sp}
		f_{|H|}(r) = \frac{4 \left(\frac{m}{\Omega_1 \Omega_2}\right)^{\frac{Nm+1}{2}}}{(Nm-1)!} r^{Nm} K_{Nm-1} \left(2 \sqrt{\frac{m}{\Omega_1 \Omega_2}}r\right).
	\end{equation}
\end{proposition}
\begin{IEEEproof}
	Without loss of generality, since in \eqref{PDF} $m_1$ and $m_2$ are used interchangeably, we set $m_1=1$ and $m_2=m$ and \eqref{PDF_sp} is derived.
\end{IEEEproof}
\begin{remark}
	From \eqref{PDF_sp}, it can be observed that the distribution of the amplitude of the sum of Rayleigh-Nakagami-$m$ random vectors is a double-Nakagami-$m$ distribution with PDF given by \eqref{dNak_PDF} where $m_1=1$, $m_2=Nm$, and replacing $\Omega_2$ with $N\Omega_2$.
\end{remark}
In the following theorem, the CDF of the amplitude of the sum of double-Nakagami-$m$ random vectors is derived in closed-form.
\begin{theorem}
	The CDF of $|H|$ can be expressed as
	\begin{equation} \label{CDF}
	\begin{split}
	F_{|H|}&(r) = \sum_{k_1=0}^{m_1-1} ... \sum_{k_N=0}^{m_1-1} \prod_{i=1}^{N} \frac{(m_2)_{m_1-1-k_i}(1-m_2)_{k_i}}{(m_1-1-k)! k_i!} \\
	& \times \left( 1 - \frac{2}{(u-1)!} \left(\sqrt{\frac{m_1 m_2}{\Omega_1 \Omega_2}} r\right)^u K_{u} \left(2\sqrt{\frac{m_1 m_2}{\Omega_1 \Omega_2}}r\right) \right).
	\end{split}
	\end{equation}
\end{theorem}
\begin{IEEEproof}
	The CDF of $|H|$ is defined as
	\begin{equation}
		F_{|H|}(r) = \int_{0}^{r} f_{|H|}(x) dx.
	\end{equation}
	Also, from \cite[03.04.21.0010.01]{wolfram}, it stands that 
	\begin{equation}
	\int x^{u} K_{u-1}(x) dx = - x^{u} K_{u}(x).
	\end{equation}
	Since $x^{u} K_{u}(x)$ is not defined in $0$, we utilize the following limit, which we prove in Appendix \ref{limit},
	\begin{equation} \label{lim}
		\lim_{x \rightarrow 0} x^v K_v(x) = 2^{v-1} (v-1)!, \quad v \in \mathbb{Z}, v > 0
	\end{equation}
	 and after some algebraic manipulations, \eqref{CDF} is derived.
\end{IEEEproof}
Next, we provide the CDF of $|H|$ for the considered special case.
\begin{proposition} 
	The CDF of $|H|$, when $H$ follows the distribution of the sum of Rayleigh-Nakagami-$m$ random vectors, can be expressed in closed-form as
	\begin{equation} \label{CDF_sp}
		\begin{split}
			F_{|H|}(r) & = 1 - \frac{2}{(Nm-1)!} \left(\sqrt{\frac{m}{\Omega_1 \Omega_2}} r\right)^{Nm} \\
				& \times K_{Nm} \left(2\sqrt{\frac{m}{\Omega_1 \Omega_2}}r\right).
		\end{split}
	\end{equation}
\end{proposition}
\begin{IEEEproof}
	Setting $m_1=1$ and $m_2=m$ in \eqref{CDF}, \eqref{CDF_sp} is derived.
\end{IEEEproof}

Next, we derive closed-form expression for the moments of the amplitude of the sum of double-Nakagami-$m$ random vectors.
\begin{theorem}
	The $n$-th moment of $|H|$ can be formulated as
	\begin{equation} \label{mom}
	\begin{split}
	\mu^n & = \left(\frac{\Omega_1 \Omega_2}{m_1 m_2}\right)^\frac{n}{2} \sum_{k_1=0}^{m_1-1} ... \sum_{k_N=0}^{m_1-1} \\
	& \prod_{i=1}^{N} \frac{(m_2)_{m_1-1-k_i}(1-m_2)_{k_i}}{(m_1-1-k)! k_i!}  \frac{\Gamma\left(\frac{n}{2}+1\right) \Gamma\left(\frac{n}{2}+u\right)}{\Gamma\left(u\right)}.
	\end{split}
	\end{equation}
\end{theorem}
\begin{IEEEproof}
	The $n$-th moment of $|H|$ is defined as
	\begin{equation}
		\mu^n = \int_{0}^{\infty} x^n f_{|H|}(x) dx.
	\end{equation}
	Using \cite[6.561/16]{Gradshteyn2014} and after some algebraic manipulations, \eqref{mom} is derived.
\end{IEEEproof}
Considering the even moments, the $l$-th moment of $|H|$ with $l=2n$ can be simplified as
\begin{equation} 
	\begin{split}
		\mu^l & = \left(\frac{\Omega_1 \Omega_2}{m_1 m_2}\right)^l \sum_{k_1=0}^{m_1-1} ... \sum_{k_N=0}^{m_1-1} \\
			& \prod_{i=1}^{N} \frac{(m_2)_{m_1-1-k_i}(1-m_2)_{k_i}}{(m_1-1-k)! k_i!} 
			l! (u)_l .
	\end{split}
\end{equation}
In the following proposition, the moments of the considered distribution are derived for the special case.
\begin{proposition} 
	The $n$-th moment of $|H|$, when $H$ follows the distribution of the sum of Rayleigh-Nakagami-$m$ random vectors, can be expressed in closed-form as
	\begin{equation} \label{mom_sp}
		\mu^n = \left(\frac{\Omega_1 \Omega_2}{m}\right)^\frac{n}{2} \frac{\Gamma\left(\frac{n}{2}+1\right) \Gamma\left(\frac{n}{2}+Nm\right)}{\Gamma\left(Nm\right)}.
	\end{equation}
\end{proposition}
\begin{IEEEproof}
	Setting $m_1=1$ and $m_2=m$ in \eqref{mom}, \eqref{mom_sp} is derived.
\end{IEEEproof}
Considering the even moments for the special case, the $l$-th moment of $|H|$ can be simplified as
\begin{equation} 
	\mu^l = \left(\frac{\Omega_1 \Omega_2}{m}\right)^l l! (Nm)_l .
\end{equation}

In the following theorem, we derive the characteristic function of the amplitude of the sum of double-Nakagami-$m$ random vectors.
\begin{theorem}
	The characteristic function of $|H|$ can be expressed in closed form as
	\begin{equation} \label{cf}
		\begin{split}
			\varphi_{|H|}(t) & = \sum_{k_1=0}^{m_1-1} ... \sum_{k_N=0}^{m_1-1} 
				\prod_{i=1}^{N} \frac{(m_2)_{m_1-1-k_i}(1-m_2)_{k_i}}{(m_1-1-k)! k_i!} \\
				& \times \left( \frac{m_1 m_2}{\Omega_1 \Omega_2} \right)^u \left( \frac{2\sqrt{m_1 m_2}}{\sqrt{\Omega_1 \Omega_2}} - j t \right)^{-2u} \frac{2^{4u}}{1+2u} \\
				& \times {}_2 \! F_1 \left( 2u , u - \frac{1}{2} , u + \frac{3}{2} , \frac{jt\sqrt{\Omega_1 \Omega_2} + 2\sqrt{m_1 m_2}}{jt\sqrt{\Omega_1 \Omega_2} - 2\sqrt{m_1 m_2}} \right),
		\end{split}
	\end{equation}
	where $t \in \mathbb{R}$ and $j^2=-1$.
\end{theorem}
\begin{IEEEproof}
	The characteristic function of $|H|$ is defined as
	\begin{equation}
		\varphi_{|H|}(t) = \int_{0}^{\infty} e^{jtx} f_{|H|}(x) dx.
	\end{equation}
	Using \cite[6.621/3]{Gradshteyn2014} and after some algebraic manipulations, \eqref{cf} is derived.
\end{IEEEproof}
In the following proposition, the characteristic function is derived for the considered special case.
\begin{proposition} 
	The characteristic function of $|H|$, when $H$ follows the distribution of the sum of Rayleigh-Nakagami-$m$ random vectors, can be expressed in closed-form as
	\begin{equation} \label{cf_sp}
		\begin{split}
			& \varphi_{|H|}(t) = \left( \frac{m_1 m_2}{\Omega_1 \Omega_2} \right)^{Nm} \left( \frac{2\sqrt{m_1 m_2}}{\sqrt{\Omega_1 \Omega_2}} - jt \right)^{-2Nm} \frac{2^{4Nm}}{1+2Nm} \\
			& \ \! \times \! {}_2 \! F_1 \! \left( 2Nm , Nm - \frac{1}{2} , Nm + \frac{3}{2} , \frac{jt\sqrt{\Omega_1 \Omega_2} + 2\sqrt{m_1 m_2}}{jt\sqrt{\Omega_1 \Omega_2} - 2\sqrt{m_1 m_2}} \right) \! .
		\end{split}
	\end{equation}
\end{proposition}
\begin{IEEEproof}
	Setting $m_1=1$ and $m_2=m$ in \eqref{cf}, \eqref{cf_sp} is derived.
\end{IEEEproof}

\section{Application in Intelligent Reflecting Surfaces}

In this section, we introduce RRSs which induce randomly selected time-variant phase shifts on the reflected signal in order to diffuse the incoming wave, investigate a practical RRS-assisted system and provide useful metrics to evaluate its performance. 

\subsection{System Model}
We consider a communication system consisting of a BS with a single antenna, an RRS and a single user with a single antenna. The transmitted signal by the BS is reflected randomly from the RRS, consisting of $N$ reflecting elements, and then is received by the user. It is assumed that the direct link from the BS to the user is blocked by obstacles, such as buildings \cite{Trigui2020}, which is a realistic assumption especially when high frequency bands are used.

The received signal, $Y$, in the user can be expressed as
\begin{equation} \label{y1}
	Y = \sqrt{l p} \sum_{k=1}^{N} e^{j \phi_k} H_{1k} H_{2k} X + W,
\end{equation}
where $X$ and $W$ denote the transmitted signal and the additive white Gaussian noise, respectively, and $e$ is the basis of the natural logarithm. The complex channel coefficients between the BS and the RRS and between the RRS and the user are denoted by $H_{1k}$ and $H_{2k}$, respectively. Moreover, $l$, $p$, and $\phi_k$ denote the equivalent path loss, the transmitted power, and the phase adjustment performed by the RRS, respectively. 
It is assumed that $|H_{1k}|$ and $|H_{2k}|$ are RVs following the Nakagami-$m$ distribution with shape parameters $m_1$ and $m_2$, respectively, and spread parameters $\Omega_1$ and $\Omega_2$, respectively, which are the same for all $N$ elements \cite{Ferreira2020}.

The $k$-th element of the RRS induces a phase shift $\phi_k$ which is uniformly distributed in $[0,2\pi]$ and is reconfigured per period equal to the coherence time.
Considering that the sum of a circular uniform RV with a RV which follows an arbitrary circular distribution results in a circular uniform RV \cite{Mardia2009}, $\theta_k = \phi_k + \arg(H_{1k}) + \arg(H_{2k})$ is uniformly distributed in $[0,2\pi]$, where $\arg(\cdot)$ denotes the argument of a complex number. 
It is obvious that the distribution of $\theta_k$ does not depend on the distribution of $\arg(H_{1k})$ and $\arg(H_{2k})$.
In this case, CSI is not necessary at the RRS which highlights the fact that the considered system is less complex than a RIS-assisted system.
It should be highlighted that if a non-reconfigurable meta-surface with predefined phase shifts is used and Rayleigh fading is considered where the phase follows the circular uniform distribution, the equivalent phase will also follow the circular uniform distribution.
However, if a meta-surface with fixed phase shifts is used in environments with different fading models, e.g., Nakagami fading, where the phase does not follow the circular uniform distribution, the signals are not diffused.
To this end, the utilization of RRSs is essential for the signals to be diffused regardless of the fading.

Furthermore, in RIS-assisted systems, the RIS attempts to cancel the phase introduced by the product of the two complex channel coefficients with the use of the phase shift $\phi_k$.
Practically, the phase correction is not perfect resulting in a phase error $\theta_k$ which is frequently modeled as a RV following the Von Mises distribution with concentration parameter $\kappa$ \cite{Badiu2020}. 
If $\kappa=0$, the distribution of the phase error $\theta_k$ is uniform and for small values of $\kappa$ the distribution is close to uniform. 
When $\kappa=0$, the phase errors are equally probable, thus the performance achieved in the RRS-assisted system can be considered as a lower bound of the performance of an RIS-assisted system.

Therefore, the received signal in \eqref{y1} can be expressed as
\begin{equation} \label{y2}
	Y = \sqrt{l p} \sum_{k=1}^{N} e^{j \theta_k} |H_{1k}| |H_{2k}| X + W.
\end{equation}

We define the composite channel coefficient $H$ as
\begin{equation} \label{channel}
	H = \sum_{k=1}^{N} e^{j \theta_k} |H_{1k}| |H_{2k}|.
\end{equation}
Therefore, the PDF of the composite channel coefficient $|H|$ is given by \eqref{PDF}, since $H$ follows the distribution of the sum of i.i.d. double-Nakagami-$m$ random vectors \cite{Ferreira2020, Trigui2020}.

\subsection{Performance Evaluation}
In this subsection, the performance of the considered system is evaluated in terms of the outage probability, the average received SNR, the ergodic capacity, the BEP, the AoF and the CQEI.

\begin{corollary}
	The outage probability can be obtained through \eqref{CDF} as
	\begin{equation} \label{Pout}
		P_{o} = F_{|H|}\left(\sqrt{\frac{\gamma_{\mathrm{thr}}}{l \gamma_t}}\right),
	\end{equation}
	where $\gamma_t$ denotes the transmitted SNR and $\gamma_{\mathrm{thr}}$ denotes the threshold which defines the outage.
\end{corollary}
If we consider the special case where one link follows the Nakagami-$m$ distribution and the other link follows the Rayleigh distribution, the outage probability can be expressed using \eqref{CDF_sp}.


In the following proposition, the average received SNR is provided and is proved to be independent of $m_1$ and $m_2$, thus it can be utilized for both considered cases.
\begin{proposition} \label{pr_SNR_r}
	The average received SNR is given by
		\begin{equation} \label{SNR_R}
			\mathbb{E}[\gamma_r] = l N \Omega_1 \Omega_2 \gamma_t.
		\end{equation}
\end{proposition}
\begin{IEEEproof}
	The average received SNR can be obtained as
	\begin{equation}
		\mathbb{E}[\gamma_r] = l \mathbb{E}\left[|H|^2\right] \gamma_t,
	\end{equation}
	where $\mathbb{E}[|H|^2]$ is the expected value of the channel gain which can be derived as the second moment in \eqref{mom}, i.e.,
	\begin{equation} \label{mom2}
		\begin{split}
			\mathbb{E}\left[|H|^2\right] & = \frac{\Omega_1 \Omega_2}{m_1 m_2} \sum_{k_1=0}^{m_1-1} ... \sum_{k_N=0}^{m_1-1} \\
				& \prod_{i=1}^{N} \frac{(m_2)_{m_1-1-k_i}(1-m_2)_{k_i}}{(m_1-1-k)! k_i!} u.
		\end{split}
	\end{equation}
	Using \cite[06.10.16.0005.01]{wolfram} and the fact that $(1)_n=n!$, $n\in\mathbb{Z},n>0$, it is proven that 
	\begin{equation} \label{sum1}
		\sum_{k_i=0}^{m_1-1} \frac{(m_2)_{m_1-1-k_i}(1-m_2)_{k_i}}{(m_1-1-k)! k_i!} = 1.
	\end{equation}
	Moreover, using \cite[06.10.17.0002.02]{wolfram}, it is proven that
	\begin{equation} \label{kPoc}
		k_i (1-m_2)_{k_i} = (1-m_2) \left((2-m_2)_{k_i} - (1-m_2)_{k_i}\right).
	\end{equation}
	Using \eqref{sum1} and \eqref{kPoc}, it can be proven that
	\begin{equation} \label{sum2}
		\sum_{k_i=0}^{m_1-1} \frac{k_i (m_2)_{m_1-1-k_i}(1-m_2)_{k_i}}{(m_1-1-k)! k_i!} = (1 - m_2) (m_1 - 1).
	\end{equation}
	Utilizing \eqref{sum1} and \eqref{sum2} and after some algebraic manipulations, \eqref{SNR_R} is derived.
\end{IEEEproof}

Next, we provide the ergodic capacity of the considered system.
\begin{proposition}
	The ergodic capacity can be expressed as
	\begin{equation} \label{ergodic}
		\begin{split}
			C & =  \frac{B}{\ln 2} \sum_{k_1=0}^{m_1-1} ... \sum_{k_N=0}^{m_1-1} 
			\prod_{i=1}^{N} \frac{(m_2)_{m_1-1-k_i}(1-m_2)_{k_i}}{(m_1-1-k)! k_i!} \\
			& \times \frac{1}{(u-1)!} G_{3,1}^{1,3} \left(\frac{l \gamma_t \Omega_1 \Omega_2}{m_1 m_2}|
				\begin{array}{c}
					1,1,1-u \\
					1 \\
				\end{array}
				\right),
		\end{split}
	\end{equation}
	where $B$ denotes the bandwidth of the fading channel.
\end{proposition}
\begin{IEEEproof}
	The ergodic capacity is calculated as the average capacity and is given by
	\begin{equation}
		C = B \int_{0}^{\infty} \log_2 \left( 1 + l \gamma_t r^2 \right) f_{|H|}(r) dr.
	\end{equation}
	Transforming $\ln(1+x)$ into Meijer's G-function \cite[01.04.26.0002.01]{wolfram} 
	and using \cite[03.04.26.0037.01]{wolfram} and \cite{Adamchik1990}, the ergodic capacity can be evaluated as
	\begin{equation} \label{ergodic_pr}
		\begin{split}
			C & = \frac{B}{\ln 2} \sum_{k_1=0}^{m_1-1} ... \sum_{k_N=0}^{m_1-1} 
				\prod_{i=1}^{N} \frac{(m_2)_{m_1-1-k_i}(1-m_2)_{k_i}}{(m_1-1-k)! k_i!} \\
				& \times \frac{1}{(u-1)!} G_{4,2}^{1,4} \left(\frac{l \gamma_t \Omega_1 \Omega_2}{m_1 m_2}|
					\begin{array}{c}
						1,1,1-u,0 \\
						1,0 \\
					\end{array}
				\right).
		\end{split}
	\end{equation}
	Using \cite[07.34.04.0002.01]{wolfram} and \cite[07.34.03.0001.01]{wolfram}, the Meijer's G-function can be simplified and \eqref{ergodic} is derived, which completes the proof.
\end{IEEEproof}
For the considered special case, the ergodic capacity can be expressed as
\begin{equation} \label{ergodic_sp}
	C =  \frac{B}{\ln 2 (Nm-1)!} G_{3,1}^{1,3} \left(\frac{l \gamma_t \Omega_1 \Omega_2}{m}|
		\begin{array}{c}
			1,1,1-Nm \\
			1 \\
		\end{array}
		\right).
\end{equation}

Next, we provide the BEP of the RRS-assisted system for binary modulations, i.e., binary phase-shift keying (BPSK), differential binary phase-shift keying (DBPSK), binary frequency-shift keying (BFSK) and noncoherent binary frequency-shift keying (NBFSK).
In this case, the BEP is given by \cite{Simon2004}
\begin{equation} \label{BEP_bin}
	P_e^b \left( H \right) = \frac{1}{2\Gamma(b)} \Gamma \left( b, a l |H|^2 \gamma_t \right),
\end{equation}
where $a$ and $b$ are modulation-dependent parameters which are presented in Table \ref{Table1} and $\Gamma(\cdot,\cdot)$ is the upper incomplete Gamma function \cite{Gradshteyn2014}.
\begin{table}
	\centering
	\caption{Values of $a$ and $b$ in \eqref{BEP_bin} for different binary modulations.}
	\begin{tabular}{|c|c|c|}
		\hline
		Modulation & $a$ & $b$ \\
		\hline 	\hline
		BPSK & $1$ & $\frac{1}{2}$ \\
		\hline
		DBPSK & $1$ & $1$ \\
		\hline
		BFSK & $\frac{1}{2}$ & $\frac{1}{2}$ \\
		\hline
		NBFSK & $\frac{1}{2}$ & $1$ \\
		\hline
	\end{tabular} 
	\label{Table1} 
\end{table}  
\begin{proposition}
	The BEP for binary modulations is given by
	\begin{equation} \label{BEP_b}
		\begin{split}
			P_e^b & = \frac{1}{2 \Gamma(b)} \sum_{k_1=0}^{m_1-1} ... \sum_{k_N=0}^{m_1-1} 
				\prod_{i=1}^{N} \frac{(m_2)_{m_1-1-k_i}(1-m_2)_{k_i}}{(m_1-1-k)! k_i!} \\
				& \times \frac{1}{(u-1)!} G_{3,2}^{2,2} \left(\frac{a l \gamma_t \Omega_1 \Omega_2}{m_1 m_2}|
				\begin{array}{c}
					1-u,0,1 \\
					0 , b \\
				\end{array}
				\right).
		\end{split}
	\end{equation}
\end{proposition}
\begin{IEEEproof}
	Transforming $\Gamma(b,x)$ into Meijer's G-function \cite[06.06.26.0005.01]{wolfram}
	and using \cite[03.04.26.0037.01]{wolfram} and \cite{Adamchik1990}, \eqref{BEP_b} is derived, which completes the proof.
\end{IEEEproof}
For the considered special case, the BEP for binary modulations can be obtained as
\begin{equation} \label{BEP_b_sp}	
	P_e^b = \frac{1}{2\Gamma(b)(Nm-1)!} G_{3,2}^{2,2} \left(\frac{a l \gamma_t \Omega_1 \Omega_2}{m}| \!
		\begin{array}{c}
			1-Nm,0,1 \\
			0 , b \\
		\end{array} \!
		\right) \! .
\end{equation}

Next, we provide the BEP of the RRS-assisted system considering $M$-ary modulations with $M \geq 4$, i.e., quadrature amplitude modulation (QAM) and PSK.
In this case, considering that Gray mapping is used, the BEP is given by \cite{Lu1999}
\begin{equation} \label{BEP_M}
	P_e \left( H \right) = a_M \sum_{k=1}^{\tau_M} \mathrm{erfc} \left( \sqrt{b_k l \gamma_t} |H| \right),
\end{equation}
where $a_M$, $\tau_M$ and $b_k$ are modulation-dependent parameters which are presented in Table \ref{Table2}.
\begin{table*}
	\centering
	\caption{Values of $\tau_M$, $a_M$ and $b_k$ in \eqref{BEP_M} for different modulations with $M \geq 4$.}
	\begin{tabular}{|c|c|c|c|}
		\hline
		Modulation & $\tau_M$ & $a_M$ & $b_k$ \\
		\hline 	\hline
		$M$-QAM & $\frac{\sqrt{M}}{2}$ & $\frac{2}{\log_2 M} \left( 1 - \frac{1}{\sqrt{M}} \right)$ & $\frac{3 \log_2 M}{2(M-1)} (2k-1)^2$ \\
		\hline
		$M$-PSK & $\max \left( \frac{M}{4} , 1 \right)$ & $\frac{1}{\max \left( \log_2 M , 2 \right)}$ & $\log_2 M \sin^2 \left( \frac{2k-1}{M} \pi \right)$ \\
		\hline
	\end{tabular} 
	\label{Table2} 
\end{table*}  
\begin{proposition}
	The BEP for $M$-ary modulations can be expressed as
	\begin{equation} \label{BEP}
	\begin{split}
	P_e & = \frac{1}{\sqrt{\pi}} \sum_{k=1}^{\tau_M} \sum_{k_1=0}^{m_1-1} ... \sum_{k_N=0}^{m_1-1} 
	\prod_{i=1}^{N} \frac{(m_2)_{m_1-1-k_i}(1-m_2)_{k_i}}{(m_1-1-k)! k_i!} \\
	& \times \frac{1}{(u-1)!} G_{3,2}^{2,2} \left(\frac{b_k l \gamma_t \Omega_1 \Omega_2}{m_1 m_2}|
	\begin{array}{c}
	1-u,0,1 \\
	0 , \frac{1}{2} \\
	\end{array}
	\right).
	\end{split}
	\end{equation}
\end{proposition}
\begin{IEEEproof}
	Transforming $\mathrm{erfc}(x)$ into Meijer's G-function \cite[06.27.26.0006.01]{wolfram}
	and using \cite[03.04.26.0037.01]{wolfram} and \cite{Adamchik1990}, \eqref{BEP} is derived, which completes the proof.
\end{IEEEproof}
For the considered special case, the BEP for $M$-ary modulations can be obtained as
\begin{equation} \label{BEP_sp}
	P_e = \frac{1}{\sqrt{\pi}(Nm-1)!} \sum_{k=1}^{\tau_M} G_{3,2}^{2,2} \left(\frac{b_k l \gamma_t \Omega_1 \Omega_2}{m}| \!
		\begin{array}{c}
			1-Nm,0,1 \\
			0 , \frac{1}{2} \\
		\end{array} \!
		\right) \! .
\end{equation}

Next, we present the AoF, a useful performance metric for the analysis of wireless communication systems which expresses the severity of the fading channel and is defined as the ratio of the variance to the square average of the instantaneous received SNR, i.e., \cite{Simon2005, Badarneh2020}
\begin{equation}
	\mathrm{AoF} = \frac{\mathbb{E}\left[ \gamma_r^2 \right] - \left( \mathbb{E}\left[ \gamma_r \right] \right)^2}{\left( \mathbb{E}\left[ \gamma_r \right] \right)^2}.
\end{equation}
In the considered RRS-assisted system, the AoF is calculated in the following proposition
\begin{proposition} \label{prop_AoF}
	The AoF is given by
	\begin{equation} \label{AoF}
		\mathrm{AoF} = 1 + \frac{1+m_1+m_2-m_1 m_2}{N m_1 m_2} .
	\end{equation}
\end{proposition}
\begin{IEEEproof}
	The proof is provided in Appendix \ref{proof_AoF}.
\end{IEEEproof}
For the considered special case, \eqref{AoF} can be further simplified as
\begin{equation}
	\mathrm{AoF} = \frac{2+Nm}{N m}.
\end{equation}

Next, we provide another useful performance metric, the CQEI, which assesses the error performance of a communication system efficiently and is defined as the ratio of the variance to the cubed mean of the instantaneous received SNR, i.e., \cite{Lioumpas2007}
\begin{equation} \label{CQEI_def}
	\mathrm{CQEI} = \frac{\mathbb{E}\left[ \gamma_r^2 \right] - \left( \mathbb{E}\left[ \gamma_r \right] \right)^2}{\left( \mathbb{E}\left[ \gamma_r \right] \right)^3}.
\end{equation}
In the following proposition, the CQEI is extracted.
\begin{proposition}
	The CQEI can be expressed as
	\begin{equation} \label{CQEI}
		\mathrm{CQEI} = \frac{1+m_1+m_2+m_1 m_2(N-1)}{N^2 m_1 m_2 \Omega_1 \Omega_2 l \gamma_t}.
	\end{equation}
\end{proposition}
\begin{IEEEproof}
	Considering that \eqref{CQEI_def} can be written as
	\begin{equation} \label{CQEI_pr}
		\mathrm{CQEI} = \frac{\mathrm{AoF}}{\mathbb{E}\left[ \gamma_r \right]},
	\end{equation}
	$\mathrm{CQEI}$ can be directly derived from \eqref{SNR_R} and \eqref{AoF}.
\end{IEEEproof}
For the considered special case, \eqref{CQEI} can be further simplified as
\begin{equation}
	\mathrm{CQEI} = \frac{2+Nm}{N^2 m \Omega_1 \Omega_2 l \gamma_t}.
\end{equation}

\subsection{Numerical Results and Discussion}
In this subsection, we illustrate the performance of the considered system.
Assuming that both the BS and the user are located at the far field region of the RRS, the equivalent path loss $l$ is given by
$l= l_1 l_2$, where $l_1$ and $l_2$ denote the path loss of the link between the BS and the RRS and the link between the RRS and the user, respectively. The path loss of each link is modeled as \cite{Wu2019}
\begin{equation}
	l_i = C_0 \left( \frac{d_i}{d_0} \right)^{-\alpha_i},
\end{equation}
where $C_0$ denotes the reference path loss at the reference distance $d_0$, $d_i$, $i\in\{1,2\}$ denotes the distance of the $i$-th link and $\alpha_i$ denotes the path loss exponent of the $i$-th link.
We set $m_1 = 3$, $m_2 = 1$, $\Omega_1=1$ and $\Omega_2=1$, which corresponds to a scenario where there is a LoS link between the BS and the RRS and a NLoS link between the RRS and the user. This scenario is motivated by the mobility of the user and, thus, the difficulty of establishing a LoS link \cite{Qian2020}. Moreover, we also examine the setup where $m_1 = 1$ and $m_2 = 1$ in order to illustrate a performance lower bound which represents a scenario where the RRS is randomly deployed, since there is no LoS link. For both path loss links, we set $C_0 = -30$dB, $d_0 = 1$m, and the path loss exponent for the link between the BS and the RRS and the link between the RRS and the user is set as $\alpha_1 = 2.8$ and $\alpha_2 = 2.2$, respectively \cite{Wu2019}.
Furthermore, unless stated otherwise, the transmitted SNR is $110$dB and it is assumed that the sum of the two distances is constant, i.e., $d_1+d_2=d$ with $d=30$m.
It is also assumed that as the number of elements increases, the size of the RRS also increases due to the fact that the inter-distance between them remains unchanged.

Fig. \ref{fig:outage} illustrates the impact of the ratio of the transmitted SNR to the outage threshold in the receiver and the number of elements of the RRS on the outage probability.
In this figure, we set the distance between the BS and the RRS as $d_1=25$m and the distance between the RRS and the user is set as $d_2=5$m.
It is observed that as the number of elements increases the outage probability decreases, thus it is clear that the QoS can be improved by increasing the number of elements without increasing the power consumption of the transceiver. Specifically, it can be observed that by doubling the number of elements the same outage performance can be achieved with $3$dB smaller transmitted SNR. However, in order to significantly improve the outage probability, the size of the RRS must be noticeably large, as in this way more electromagnetic power density will impinge upon the surface and, thus, more energy will be focused on the user leading to an increase of the received SNR.

\begin{figure}
	\centering
	\begin{tikzpicture}
	\begin{semilogyaxis}[
	width=0.9\linewidth,
	xlabel = {$\frac{\gamma_t}{\gamma_{\mathrm{thr}}}$ (dB)},
	ylabel = {$P_{o}$},
	xmin = 80,xmax = 120,
	ymin = 0.001,
	ymax = 1,
	grid = major,
	legend entries = {{$N=32$},{$N=64$},{$N=128$},{$N=256$}},
	legend cell align = {left},
	legend pos = south west
	]
	\addplot[
	black,
	mark=square,
	mark repeat = 1,
	mark size = 2,
	line width = 1pt,
	style = solid,
	]
	table {data/outage/out32.dat};
	\addplot[
	black,
	mark=diamond,
	mark repeat = 1,
	mark size = 2,
	line width = 1pt,
	style = solid,
	]
	table {data/outage/out64.dat};
	\addplot[
	black,
	mark=triangle,
	mark repeat = 1,
	mark size = 2,
	line width = 1pt,
	style = solid,
	]
	table {data/outage/out128.dat};
	\addplot[
	black,
	mark=o,
	mark options={solid},
	mark repeat = 1,
	mark size = 2,
	line width = 1pt,
	style = solid,
	]
	table {data/outage/out256.dat};	
	\end{semilogyaxis}
	\end{tikzpicture}
	\caption{Outage probability $P_{o}$ versus $\frac{\gamma_t}{\gamma_{\mathrm{thr}}}$.}
	\label{fig:outage}
\end{figure}
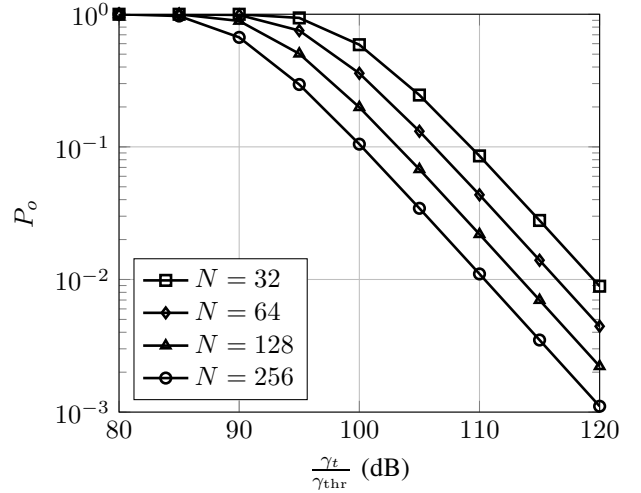

\begin{figure}
	\centering
	\begin{tikzpicture}
	\begin{axis}[
	width=0.9\linewidth,
	xlabel = {$w$} ,
	ylabel = {$\mathbb{E}[\gamma_r]$ (dB)},
	xmin = 0.1,xmax = 0.9,
    ymin=0,
	ymax = 30,
	xtick = {0.1,0.3,...,0.9},
	grid = major,
	legend entries = {{$N=32$},{$N=64$},{$N=128$},{$N=256$}},
	legend cell align = {left},
	legend style={at={(0.32,1)},anchor=north west}
	]
	\addplot[
	black,
	mark=square,
	mark repeat = 2,
	mark size = 2,
	line width = 1pt,
	style = solid,
	]
	table {data/average_snr/y1.dat};
	\addplot[
	black,
	mark=diamond,
	mark repeat = 2,
	mark size = 2,
	line width = 1pt,
	style = solid,
	]
	table {data/average_snr/y2.dat};
	\addplot[
	black,
	mark=triangle,
	mark repeat = 2,
	mark size = 2,
	line width = 1pt,
	style = solid,
	]
	table {data/average_snr/y3.dat};
	\addplot[
	black,
	mark=o,
	mark repeat = 2,
	mark size = 2,
	line width = 1pt,
	style = solid,
	]
	table {data/average_snr/y4.dat};
	\end{axis}
	\end{tikzpicture}
	\caption{Average received SNR versus $w$.}
	\label{fig:averageSNR}
\end{figure}
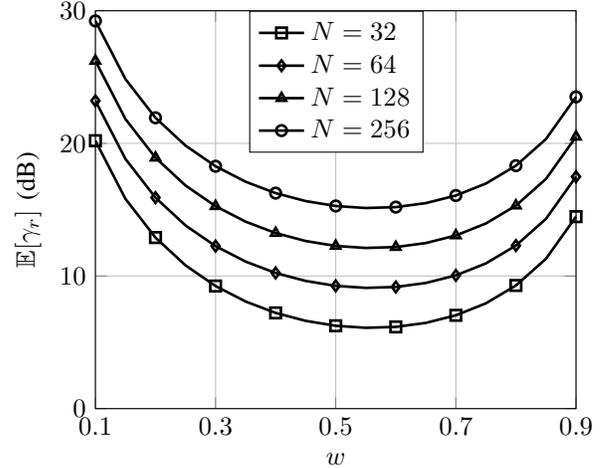

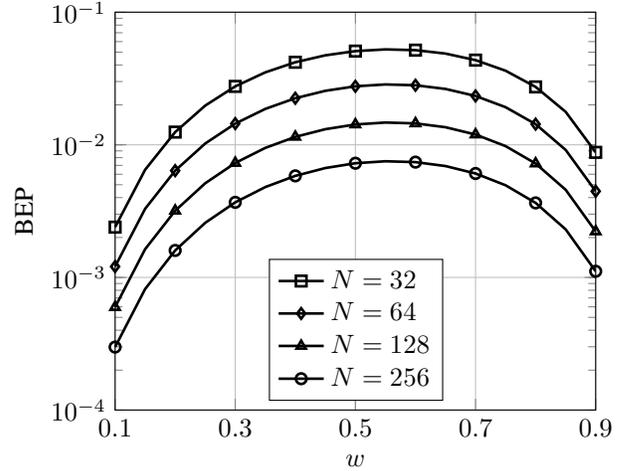
\begin{figure}
	\centering
	\begin{tikzpicture}
	\begin{semilogyaxis}[
	width=0.9\linewidth,
	xlabel = {$w$} ,
	ylabel = {BEP},
	xmin = 0.1,xmax = 0.9,
    ymin=0.0001,
	ymax = 0.1,
	xtick = {0.1,0.3,...,0.9},
	grid = major,
	legend entries = {{$N=32$},{$N=64$},{$N=128$},{$N=256$}},
	legend cell align = {left},
	legend style={at={(0.32,0.38)},anchor=north west}
	]
	\addplot[
	black,
	mark=square,
	mark repeat = 2,
	mark size = 2,
	line width = 1pt,
	style = solid,
	]
	table {data/bep/bep32.dat};
	\addplot[
	black,
	mark=diamond,
	mark repeat = 2,
	mark size = 2,
	line width = 1pt,
	style = solid,
	]
	table {data/bep/bep64.dat};
	\addplot[
	black,
	mark=triangle,
	mark repeat = 2,
	mark size = 2,
	line width = 1pt,
	style = solid,
	]
	table {data/bep/bep128.dat};
	\addplot[
	black,
	mark=o,
	mark options={solid},
	mark repeat = 2,
	mark size = 2,
	line width = 1pt,
	style = solid,
	]
	table {data/bep/bep256.dat};
	\end{semilogyaxis}
	\end{tikzpicture}
	\caption{BEP versus $w$ for BPSK or 4-QAM modulation with Gray mapping.}
	\label{fig:BEP}
\end{figure}

In Fig. \ref{fig:averageSNR} and Fig. \ref{fig:BEP}, we set $d_1=wd$ and $d_2 = (1-w)d$, where $w \in [0.1,0.9]$ to ensure that the terminals are located at the far field region, and the average received SNR and the BEP versus $w$ are illustrated, respectively. For the BEP, the modulation is assumed to be either BPSK or 4-QAM with Gray mapping. 
It is observed that as the number of elements increases, both the average received SNR and the BEP increases at a certain distance.
Moreover, it should be highlighted that the placement of the RRS plays an important role and it becomes evident that it should be placed either close to the BS or close to the user, since the path losses are maximized when the RRS is placed in an intermediate point between the BS and the user. Also, the fact that the performance is slightly improved when the RRS is placed close to the BS is justified as the exponent $a_1$ is larger than $a_2$.

\begin{figure}
	\centering
	\begin{tikzpicture}
	\begin{axis}[
	width=0.9\linewidth,
	xlabel = {$\gamma_t$ (dB)} ,
	ylabel = {$C/B$ (bps/Hz)},
	xmin = 80,xmax = 120,
    ymin=0,
	ymax = 10,
	grid = major,
	legend entries = {{$N=32$},{$N=64$},{$N=128$},{$N=256$}},
	legend cell align = {left},
	legend pos = north west,
	]
	\addplot[
	black,
	mark=square,
	mark repeat = 1,
	mark size = 2,
	line width = 1pt,
	style = solid,
	]
	table {data/ec/ec32.dat};
	\addplot[
	black,
	mark=diamond,
	mark repeat = 1,
	mark size = 2,
	line width = 1pt,
	style = solid,
	]
	table {data/ec/ec64.dat};
	\addplot[
	black,
	mark=triangle,
	mark repeat = 1,
	mark size = 2,
	line width = 1pt,
	style = solid,
	]
	table {data/ec/ec128.dat};
	\addplot[
	black,
	mark=o,
	mark options={solid},
	mark repeat = 1,
	mark size = 2,
	line width = 1pt,
	style = solid,
	]
	table {data/ec/ec256.dat};
	\end{axis}
	\end{tikzpicture}
	\caption{Normalized ergodic capacity $\frac{C}{B}$ versus $\gamma_t$.}
	\label{fig:EC}
\end{figure}
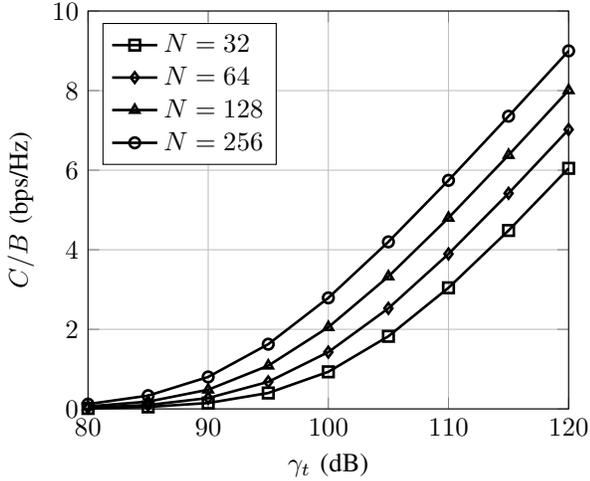

Fig. \ref{fig:EC} depicts the impact of the transmitted SNR on the  ergodic capacity for different numbers of elements of the RRS. The distances of the two links are set as in Fig. \ref{fig:outage}. It is observed that high transmitted SNR values are required for successful information transmission. Moreover, increasing the number of elements leads to better ergodic capacity, since the average received SNR is proportional to the number of elements and the QoS can be improved by the enlargement of the RRS without consuming more power.

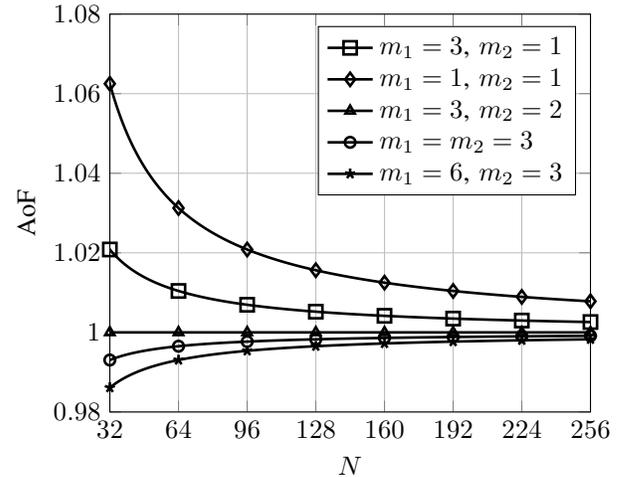
\begin{figure}
	\centering
	\begin{tikzpicture}
	\begin{axis}[
	width=0.9\linewidth,
	xlabel = {$N$} ,
	ylabel = {AoF},
	xmin = 32,xmax = 256,
    ymin=0.98,
	ymax = 1.08,
	xtick = {32,64,...,256},
	grid = major,
	legend entries = {{$m_1=3$, $m_2=1$},{$m_1=1$, $m_2=1$},{$m_1=3$, $m_2=2$},{$m_1=m_2=3$},{$m_1=6$, $m_2=3$}},
	legend cell align = {left},
	legend pos = north east,
	]
	\addplot[
	black,
	mark=square,
	mark repeat = 32,
	mark size = 2.5,
	line width = 1pt,
	style = solid,
	]
	table {data/aof/aof13.dat};
	\addplot[
	black,
	mark=diamond,
	mark repeat = 32,
	mark size = 2.5,
	line width = 1pt,
	style = solid,
	]
	table {data/aof/aof11.dat};
	\addplot[
	black,
	mark=triangle,
	mark repeat = 32,
	mark size = 2,
	line width = 1pt,
	style = solid,
	]
	table {data/aof/aof23.dat};
	\addplot[
	black,
	mark=o,
	mark repeat = 32,
	mark size = 2,
	line width = 1pt,
	style = solid,
	]
	table {data/aof/aof33.dat};
	\addplot[
	black,
	mark=star,
	mark repeat = 32,
	mark size = 2,
	line width = 1pt,
	style = solid,
	]
	table {data/aof/aof63.dat};
	\end{axis}
	\end{tikzpicture}
	\caption{AoF versus the number of elements $N$.}
	\label{fig:AoF}
\end{figure}

Fig. \ref{fig:AoF} illustrates how the AoF is affected by the number of elements of the RRS. As the number of elements increases, the AoF converges to $1$. Especially for the case where $m_1 = 3$ and $m_2 = 2$ the amount of fading equals to $1$ regardless of the number of elements. As the number of elements increases, due to the fact that phase shifts are circular uniformly distributed, more non-coherent paths are created. 
Therefore, when a large RRS with arbitrary phase shift is deployed between two terminals, the AoF tends to $1$, which coincides to the value of the AoF for a single Rayleigh channel. 
Moreover, as the AoF is hardly affected by the fading conditions for large number of elements, it follows that the fading conditions have limited impact on the performance of the considered system. It should be highlighted that as $m_1$ and $m_2$ are used interchangeably, the AoF is the same for the scenarios where the value of the shape parameters of both links is interchanged, e.g., AoF$_{3,1}$ = AoF$_{1,3}$.

\begin{figure}
	\centering
	\begin{tikzpicture}
	\begin{semilogyaxis}[
	width=0.9\linewidth,
	xlabel = {$\gamma_t$ (dB)} ,
	ylabel = {CQEI},
	xmin = 80,xmax = 120,
    ymin=0.001,
	ymax = 100,
	grid = major,
	legend entries = {{$N=32$},{$N=64$},{$N=128$},{$N=256$}},
	legend cell align = {left},
	legend pos = south west,
	]
	\addplot[
	black,
	mark=square,
	mark repeat = 1,
	mark size = 2,
	line width = 1pt,
	style = solid,
	]
	table {data/cqei/cqei1332.dat};
	\addplot[
	black,
	mark=diamond,
	mark options={solid},
	mark repeat = 1,
	mark size = 2,
	line width = 1pt,
	style = solid,
	]
	table {data/cqei/cqei1364.dat};
	\addplot[
	black,
	mark=triangle,
	mark repeat = 1,
	mark size = 2,
	line width = 1pt,
	style = solid,
	]
	table {data/cqei/cqei13128.dat};
	\addplot[
	black,
	mark=o,
	mark options={solid},
	mark repeat = 1,
	mark size = 2,
	line width = 1pt,
	style = solid,
	]
	table {data/cqei/cqei13256.dat};
	\end{semilogyaxis}
	\end{tikzpicture}
	\caption{CQEI versus $\gamma_t$.}
	\label{fig:CQEI}
\end{figure}
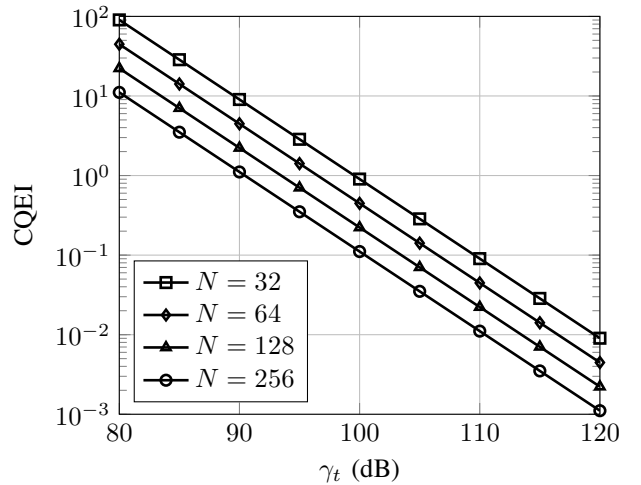

Fig. \ref{fig:CQEI} depicts the behavior of the CQEI as the transmitted SNR increases. It should be highlighted that the performance improves as the CQEI decreases. It is clear that as the number of elements increases, the CQEI decreases without requiring larger transmitted SNR and as a consequence the system performance upgrades.


\section{Conclusions}

In this work, we have introduced RRSs, being defined as reconfigurable meta-surfaces whose elements induce a randomly selected time-variant phase shift on the reflected signal in order to approximate the diffusion function of a RIS.
We have utilized the distribution of the sum of double-Nakagami-$m$ random vectors in a RRS-assisted system where the two links undergo Nakagami-$m$ fading and equivalent phases following the circular uniform distribution are considered.
We have derived closed-form expressions of useful metrics such as the outage probability, the average received SNR, the ergodic capacity, the BEP, the AoF and the CQEI.
It has been proved that the average received SNR is proportional to the number of elements $N$ and it can be observed that large number of elements is required in order to improve the system performance. Furthermore, we can enhance the system performance by deploying the RRS in the vicinity of the BS or the user.  Finally, it has been illustrated that when a large RRS with arbitrary phase shifts is deployed between two terminals, the AoF tends to $1$, which coincides to the value of the AoF for a single Rayleigh channel.

\appendices
\section{Proof of Theorem \ref{theorem}} \label{proof_theorem}
The PDF of $|H|$ is given by \cite{Abdi2000}
\begin{equation} \label{abdi}
f_{|H|}(r) = r \int_{0}^{\infty} \rho J_0 (r \rho) \Lambda(\rho) d\rho,
\end{equation}
where $J_v()$ is the $v$-th order Bessel function of the first kind \cite{Gradshteyn2014} and $\Lambda(\rho)$ can be expressed as
\begin{equation} \label{lambda}
\Lambda(\rho) = \mathbb{E}_{h_1...h_N} \left[ \prod_{i=1}^{N} J_0 (h_i \rho) \right] = \prod_{i=1}^{N} \mathbb{E}_{h_i} \left[ J_0 (h_i \rho) \right].
\end{equation}
The expected value in \eqref{lambda} can be evaluated as 
\begin{equation} \label{exp1}
\mathbb{E}_{h_i} \left[ J_0 (h_i \rho) \right] = \int_{0}^{\infty} J_0 (z \rho) f_{h_i}(z) dz,
\end{equation}
since $h_1,...,h_N$ are independent.
Transforming the Bessel functions into Meijer’s G-functions \cite[03.04.26.0037.01]{wolfram}, \cite[03.01.26.0056.01]{wolfram} and using \cite{Adamchik1990}, \eqref{exp1} can be written as 
\begin{equation} \label{exp2}
\begin{split}
\mathbb{E}_{h_i} \left[ J_0 (h_i \rho) \right] & = \frac{1}{\Gamma (m_1) \Gamma (m_2)} \\
& \times G_{2,2}^{1,2}\left(\frac{ \Omega_1 \Omega_2 \rho^2}{4 m_1 m_2}|
\begin{array}{c}
1-m_1,1-m_2 \\
0,0 \\
\end{array}
\right),
\end{split}
\end{equation}
where $G[\cdot]$ is the Meijer’s G-function \cite{Gradshteyn2014}. Using \cite[07.23.26.0004.01]{wolfram}, \eqref{exp2} can be expressed as
\begin{equation} \label{exp3}
\mathbb{E}_{h_i} \left[ J_0 (h_i \rho) \right] = \, _2F_1\left(m_1,m_2;1;-\frac{ \Omega_1 \Omega_2 \rho^2}{4 m_1 m_2}\right),
\end{equation}
where $\, _2F_1(\cdot,\cdot,\cdot,\cdot)$ is the Gauss hypergeometric function \cite{Gradshteyn2014}.
Using \cite[07.23.17.0046.01]{wolfram}, the hypergeometric function can be expressed as a finite sum and \eqref{exp3} can be rewritten as
\begin{equation} \label{exp4}
\begin{split}
\mathbb{E}_{h_i} \left[ J_0 (h_i \rho) \right] & = \frac{(m_1)_{m_2-1}}{(m_2-1)!} \sum_{k=0}^{m_1-1} \frac{(1-m_1)_{k}(1-m_2)_{k}}{(2-m_1-m_2)_{k} k!} \\
& \times \left(\frac{4 m_1 m_2}{4 m_1 m_1+\Omega_1 \Omega_2 \rho^2}\right)^{m_1+m_2-k-1}.
\end{split}
\end{equation}
Considering that all $N$ vectors have the same $m_1$, $m_2$, $\Omega_1$ and  $\Omega_2$, using \eqref{abdi} and \eqref{exp4} and after some algebraic manipulations similar to \cite{Karagiannidis2006b}, the PDF of $|H|$ is given by 
\begin{equation} \label{pdf_pr}
\begin{split}
f_{|H|}(r) & = r \left( \frac{(m_1)_{m_2-1}}{(m_2-1)!} \right)^N \sum_{k_1=0}^{m_1-1} ... \sum_{k_N=0}^{m_1-1} \left(\frac{4 m_1 m_2}{\Omega_1 \Omega_2}\right)^u \\
& \times \int_{0}^{\infty} \rho J_0 (r \rho) \left(\frac{1}{\frac{4 m_1 m_2}{\Omega_1 \Omega_2}+\rho^2}\right)^u d\rho.
\end{split}
\end{equation}
It should be highlighted that $u\in\mathbb{Z},u>0$, since $m_1,m_2 \in \mathbb{Z}$.
Utilizing \cite[6.565/4]{Gradshteyn2014} in \eqref{pdf_pr} and after some algebraic manipulations, \eqref{PDF} is derived and  the proof is completed.

\section{Proof of \eqref{lim}} \label{limit}
	Using L'Hospital's rule and \cite[03.04.20.0005.01]{wolfram}, it stands that
	\begin{equation}
	\lim_{x \rightarrow 0}  \frac{K_v(x)}{x^{-v}} = -\lim_{x \rightarrow 0}  \frac{K_v(x)}{x^{-v}} + \lim_{x \rightarrow 0}  \frac{x^{v+1}}{v}K_{v+1}(x).
	\end{equation}
	Thus, the following expression stands
	\begin{equation} \label{lim_pr1}
	2v \lim_{x \rightarrow 0} x^{v}K_{v}(x) = \lim_{x \rightarrow 0} x^{v+1}K_{v+1}(x).
	\end{equation}
	Using \eqref{lim_pr1} and considering that $v \in \mathbb{Z}, v > 0$, the following expression can be derived
	\begin{equation} \label{lim_pr2}
	\lim_{x \rightarrow 0} x^v K_v(x) = 2^{v-1} (v-1)! \lim_{x \rightarrow 0} x K_1(x) .
	\end{equation}
	An approximation of $K_v(x)$ as $x\rightarrow 0$, if $\mathrm{Re}\{v\} > 0$, where $\mathrm{Re}\{\cdot\}$ denotes the real part of a complex number, is given by \cite{Abramowitz1970}
	\begin{equation}
	K_v(x) \simeq \frac{1}{2} \Gamma(v) \left( \frac{1}{2} x \right)^{-v}.
	\end{equation}
	Hence, the following limit can be derived
	\begin{equation} \label{lim_pr3}
	\lim_{x \rightarrow 0} x K_1(x) = 1.
	\end{equation}
	Utilizing \eqref{lim_pr2} and \eqref{lim_pr3}, the proof is completed.

\section{Proof of Proposition \ref{prop_AoF}}	\label{proof_AoF}
	To prove \eqref{AoF}, $\mathbb{E}\left[ \gamma_r^2 \right]$ should be calculated, since $\mathbb{E}\left[ \gamma_r \right]$ can be obtained from \eqref{SNR_R}.
	The second moment of the received SNR can be obtained as
	\begin{equation}
	\mathbb{E}[\gamma_r^2] = l^2 \mathbb{E}\left[|H|^4\right] \gamma_t^2,
	\end{equation}
	where $\mathbb{E}[|H|^4]$ can be derived as the fourth moment in \eqref{mom}, i.e.,
	\begin{equation} \label{mom4}
	\begin{split}
	\mathbb{E}\left[|H|^4\right] & = 2\left(\frac{\Omega_1 \Omega_2}{m_1 m_2}\right)^2 \sum_{k_1=0}^{m_1-1} ... \sum_{k_N=0}^{m_1-1} \\
	& \prod_{i=1}^{N} \frac{(m_2)_{m_1-1-k_i}(1-m_2)_{k_i}}{(m_1-1-k)! k_i!} u(u+1).
	\end{split}
	\end{equation}
	Utilizing the proof of Proposition \ref{pr_SNR_r}, \eqref{mom4} can be written as
	\begin{equation} \label{mom4_2}
	\begin{split}
	\mathbb{E}\left[|H|^4\right] & = 2\left(\frac{\Omega_1 \Omega_2}{m_1 m_2}\right)^2 \sum_{k_1=0}^{m_1-1} ... \sum_{k_N=0}^{m_1-1} \\
	& \prod_{i=1}^{N} \frac{(m_2)_{m_1-1-k_i}(1-m_2)_{k_i}}{(m_1-1-k)! k_i!} u^2 - 2 N \frac{\Omega_1^2 \Omega_2^2}{m_1 m_2}.
	\end{split}
	\end{equation}
	The first term in \eqref{mom4_2}, termed as $T$, considering the definition of $u$ can be written as
	\begin{equation} \label{mom4_3}
	\begin{split}
	T_1 & = 2 \sum_{k_1=0}^{m_1-1} ... \sum_{k_N=0}^{m_1-1} \prod_{i=1}^{N} \frac{(m_2)_{m_1-1-k_i}(1-m_2)_{k_i}}{(m_1-1-k)! k_i!} \\
	& \times \left( \left(N(m_1+m_2-1)\right)^2 - 2N(m_1+m_2-1) \sum_{i=1}^N k_i \right. \\ 
	& \left. + \left(\sum_{i=1}^N k_i\right)^2 \right) \left(\frac{\Omega_1 \Omega_2}{m_1 m_2}\right)^2 .
	\end{split}
	\end{equation}
	The first and the second term in \eqref{mom4_3} can be calculated utilizing the proof of Proposition \ref{pr_SNR_r}. 
	Using \cite[06.10.17.0002.02]{wolfram} and \eqref{kPoc}, it is proven that
	\begin{equation} \label{kkPoc}
	\begin{split}
	k_i^2 (1-m_2)_{k_i} & = (1-m_2) \left( (2-m_2) (3-m_2)_{k_i} \right. \\
	& \left. - (3-2m_2) (2-m_2)_{k_i} + (1-m_2)_{k_i}\right).
	\end{split}
	\end{equation}
	Using \eqref{sum1}, \eqref{sum2} and \eqref{kkPoc}, it can be proven that
	\begin{equation} \label{sum3}
	\begin{split}
	& \sum_{k_i=0}^{m_1-1} \frac{k_i^2 (m_2)_{m_1-1-k_i}(1-m_2)_{k_i}}{(m_1-1-k)! k_i!} = (1 - m_2) \\
	& \quad \times \left( (2-m_2)\frac{m_1(m_1+1)}{2} - (3-2m_2)m_1 + 1-m_2 \right).
	\end{split}
	\end{equation}
	Utilizing \eqref{sum1}, \eqref{sum2} and \eqref{sum3} and after some algebraic manipulations, the third term in \eqref{mom4_3}, termed as $T_2$, can be calculated as
	\begin{equation} \label{mom4_4}
	\begin{split}
	T_2 & = 2\left(\frac{\Omega_1 \Omega_2}{m_1 m_2}\right)^2 N \left( \frac{(m_1-1)(m_2-1)}{2} (2 - 2m_2 \right. \\
	& \left. + m_1(m_2-2)) + (N-1) ((1-m_2)(m_1-1))^2 \right). 
	\end{split}
	\end{equation}
	Using \eqref{mom4_4} and after some algebraic manipulations, \eqref{AoF} can be derived which completes the proof.


\bibliographystyle{IEEEtran}
\bibliography{Bibliography}


\end{document}